\documentclass{article}
\usepackage[preprint]{log_2022}
\usepackage[numbers,compress,sort]{natbib}
\usepackage[backref=page]{hyperref}
\usepackage{booktabs,multirow,amsfonts,graphicx,cleveref,dcolumn}
\usepackage{array}
\usepackage{subcaption}
\usepackage{float}
\usepackage{xcolor}

\newcommand{\inl}{\text{inlet}}
\newcommand{\outl}{\text{outlet}}

\newcommand{\resqm}{R_{\text{quad}}}
\newcommand{\reslm}{R_{\text{lin}}}
\newcommand{\rri}{\text{RRI}}

\usepackage{comments}
\Crefname{equation}{Eq.}{Eqs.}

\title[Hybrid Physics-Based and Data-Driven Modeling of Vascular Bifurcation Pressure Differences]{Hybrid Physics-Based and Data-Driven Modeling of Vascular Bifurcation Pressure Differences}

\author[Rubio et al.]{%
Natalia L. Rubio\\
\institute{Stanford University}\\\
\email{nrubio@stanford.edu}\\\And
Luca Pegolotti\\
\institute{Stanford University}\\\And
Martin R. Pfaller\\
\institute{Stanford University}\\\And
Eric F. Darve\\
\institute{Stanford University}\\\And
Alison L. Marsden\\
\institute{Stanford University}\\
\email{amarsden@stanford.edu}
}

\begin{document}
\maketitle

\begin{abstract}
Reduced-order models (ROMs) allow for the simulation of blood flow in patient-specific vasculatures without the high computational cost and wait time associated with traditional computational fluid dynamics (CFD) models.  Unfortunately, due to the simplifications made in their formulations, ROMs can suffer from significantly reduced accuracy.  One common simplifying assumption is the continuity of static or total pressure over vascular junctions.  In many cases, this assumption has been shown to introduce significant error.  We propose a model to account for this pressure difference, with the ultimate goal of increasing the accuracy of cardiovascular ROMs.  Our model successfully uses a structure common in existing ROMs in conjunction with machine-learning techniques to predict the pressure difference over a vascular bifurcation. We analyze the performance of our model on steady and transient flows, testing it on three bifurcation cohorts representing three different bifurcation geometric types.  We also compare the efficacy of different machine-learning techniques and two different model modalities.
\end{abstract}

\section{Introduction}

In the last 20 years, computational fluid dynamics (CFD) simulations of cardiovascular flows have been established as a valuable tool in cardiovascular medicine, science, and engineering \cite{Schwarz2023BeyondDisease, figueroa17Blood, Kamada2022BloodDiseases, Morris201ComputationalMedicine, Lee2011ComputationalDisease}.  First, they shed insight into the clinical management of cardiovascular disease.  Cardiovascular flow simulations are used to analyze the flows in patient-specific vasculatures and the associated health outcomes.  For instance, in \cite{Sengupta2012Image-basedDisease, Sengupta2014ThromboticDisease, Menon2022TheHemodynamics, GrandeGutierrez2019HemodynamicDisease}, CFD models were used to analyze flow through coronary artery aneurysms of patients with Kawasaki disease, and the analysis yielded metrics that correlated with thrombotic risk. CFD simulations are also used to inform surgical planning.  They provide patient-specific insights into the characteristics of a diseased anatomy that allow clinicians to customize the patient's treatment, rather than using a ``one size fits all'' approach.  Furthermore, CFD models predict the changes in flow behavior that would result from hypothetical surgical modifications.  For instance, in the context of multi-stage surgical intervention for single-ventricle heart defects, CFD flow analysis was used to compare two candidate Stage 2 operations, the hemi-Fontan and bidirectional Glenn procedures, study patient-specific effects of the hemi-Fontan procedure under varying physiological states, and analyze the effects of geometric variations in anatomies constructed by the Stage 3 Fontan procedure \cite{Migliavacca2003ComputationalAnastomosis, Bove2003ComputationalSyndrome, Kung2013PredictiveCases}. Second, cardiovascular flow simulations have also contributed to our understanding of cardiovascular biomechanics by characterizing hemodynamics that are difficult to observe experimentally.  For example, CFD simulations made instrumental contributions to our understanding of the mechanisms driving the development of pulmonary hypertension by modeling flows in vessels that are difficult to measure experimentally \cite{Dong2021ComputationalDefects, Hunter2008PulmonaryHypertension, Yang2019EvolutionPatients, Tang2012WallStudy}.  Third, cardiovascular flow simulations allow for design testing in a low-cost, low-risk setting, which is invaluable in the engineering and optimization of cardiovascular medical devices and treatments including stents \cite{Valentim2023SystematicStents, Gundert2012OptimizationDynamics,Frank2002ComputationalDesign}, grafts \cite{Sankaran2012Patient-specificSurgery, Seo2022ComputationalSurgery, Ramachandra2016Patient-SpecificGrafts, Yang2010ConstrainedExercise}, circulatory support systems \cite{Yang2023PassiveSupport, Fraser2011TheDevices, Bluestein2017UtilizingClinic, Engelke2018CompetingStudy}, and vascular drug delivery systems \cite{hossain12mathematical, bao14USNCTAM, Meschi2021FlowDisease}.

Traditional CFD models solve the unsteady Navier-Stokes equations in three dimensions (3D), a computationally intensive task with limited practicality in clinical settings, real-time analysis, and multi-query applications.  Reduced-order models (ROMs) are simplified representations of cardiovascular flows that predict bulk properties at much lower computational cost, providing a computationally tractable alternative to 3D simulations.  ROMs have been used to find boundary and initial conditions for higher-fidelity simulations \cite{Pfaller2021OnSimulations, Nair2023Non-invasiveMeasurements, Kim2010Patient-specificArteries, GrandeGutierrez2022AHemodynamics, Brown2023AMechanics, Olufsen1999StructuredArteries}, for uncertainty quantification \cite{Seo2020TheWalls, Seo2020Multi-FidelityUncertainty, Tran2017AutomatedSimulations}, and as stand-alone models \cite{Muller2014ASystem, Zhang2016DevelopmentData, Pham2023SvMorph:Anatomies}.  One-dimensional (1D) and zero-dimensional (0D) ROMs are two of the most frequently used ROMs in the cardiovascular flow modeling community \cite{Hughes1973OnVessels, Stergiopulos1992ComputerA, Westerhof1983CoronaryWaterfall, Shi2011ReviewSystem, Peiro2009ReducedSystem}.

In the 1D ROM, the flow through the vasculature is modeled using a one-dimensional PDE enforcing conservation of mass and momentum, derived by integrating across the vessel cross-section \cite{Ghigo2018AModel, Canic2003MathematicalVessels, Hughes1973OnVessels, Olufsen20045.Arteries, Wan2002ADisease, Wang2015VerificationModel, Wang2016FluidModel}.  The 0D ROM is a formulation in which the vasculature is represented by an idealized electric circuit in which flow and pressure are analogous to current and voltage, respectively. Vessels are represented by wires containing circuit elements (e.g., resistors, capacitors, and inductors) with characteristic values capturing the 3D vessel geometry.  The bulk flow and pressure values at the inlets and outlets of each branch in the vasculature are given by the values of current and pressure at the corresponding nodes in the analogous electric circuit \cite{Kim2010Patient-specificArteries, Pfaller2022AutomatedFlow, Mirramezani2019ReducedArteries, Formaggia2010CardiovascularSystem}.

While ROMs are promising tools for high-speed, computationally lightweight modeling of cardiovascular flows, they suffer from reduced accuracy due to the simplifications in their formulations.  One such simplification occurs in the handling of bifurcations.  Bifurcations generally feature flow separation and other nonlinear behaviors that cannot be modeled with current ROMs. Frequently, continuity of either static \cite{Taylor-LaPole2023APatients, Olufsen1999StructuredArteries, Stergiopulos1992ComputerA, Reymond2009ValidationTree, Mynard2016NovelPower} or total \cite{Fullana2009ANetworks, Sherwin2003ComputationalSystem, Lee2016MultiphysicsboldsymbolmathcalCmathbfHeart, Alastruey2009ModellingSynthesis, Alastruey2012ArterialHaemodynamics, Mynard2008AMethod}  pressure is assumed over a bifurcation. Practical experience and high-fidelity CFD simulations, however, indicate that significant differences in both static and total pressure between the inlets and outlets of a bifurcation can exist.  Other researchers have also found that the treatment of bifurcations has a significant effect on ROM solutions \cite{Nair2023HemodynamicsSimulation, San2012AnFlows, Pfaller2022AutomatedFlow, Chnafa2017ImprovedDrops, Steele2003InGrafts,  Huberts2012AClinic, Wood1993ModelingJunctions}.  As such, there is a need to develop and incorporate models that accurately predict pressure differences over vascular bifurcations to improve ROM accuracy.  Previous researchers have proposed such models specifically for cardiovascular flows and for other flow networks featuring bifurcations \cite{Gardel1971LesConique, Bassett2003ASimulations, Mynard2015AJunctions}.  These models vary in complexity and are generally developed using a combination of physics-based and empirical approaches.  

In recent years, as computational blood flow models have gained traction, more cardiovascular CFD data has become available \cite{Wilson2013TheResults}.  In parallel, significant advances have been made in data-driven modeling techniques.  Given these new tools, we present a novel, hybrid approach for modeling pressure differences over bifurcations in ROMs.  Specifically, we propose to account for pressure differences by augmenting bifurcation outlet branches with a 0D bifurcation element comprising a serially-connected resistor, quadratic resistor, and inductor whose characteristic values are determined from the bifurcation geometry using machine learning (ML) techniques.  In doing so, we apply ML techniques to our problem within a constrained, physics-based form that reduces the complexity of the regression problem and provides interpretability.

We first discuss the structure of our model and how it is intended to integrate with ROM frameworks.  Next, we describe the procedure used to generate training data for the machine learning components of our model.  We show that our hybrid physics-based data-driven model structure is able to accurately predict the pressure difference over bifurcations with previously unseen geometries for three different cohorts of geometries in both steady and transient flow settings.  We consider several ML regression techniques and compare their effectiveness for our task.  Finally, we discuss the contributions of our model towards reduced-order cardiovascular modeling, summarize its current limitations, and propose future work to develop the model further and deploy it in existing ROMs.

\section{Methods}

\begin{figure}[htbp]
\centering
{\includegraphics[width=0.95 \textwidth]{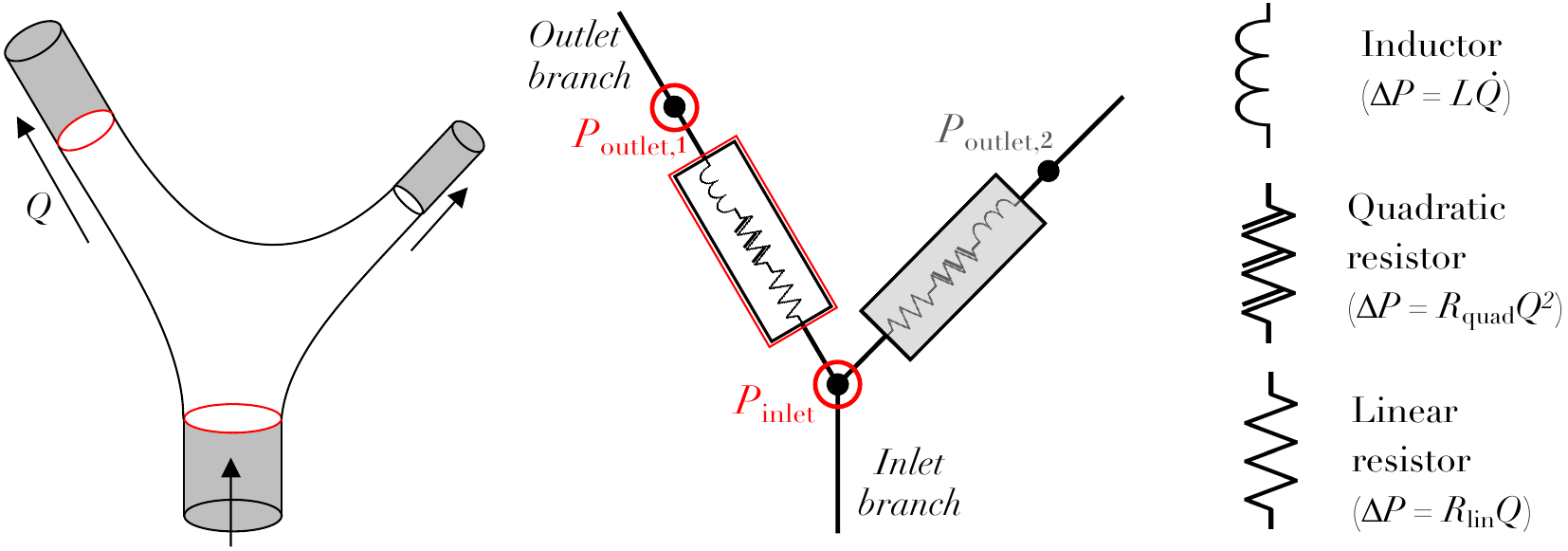}}
\caption{3D vascular bifurcation (left) and its representation in a ROM (center) including the proposed RRI bifurcation block, featuring a linear resistor, quadratic resistor, and inductor (right).}
\label{fig:3d_to_0d}
\end{figure}

In most ROM solvers, and for the purposes of this work, pressure drops are computed between the inlet and each outlet individually.  We indicate the bifurcation inlet with the subscript \textit{``inlet''}, the outlet over which we are currently computing the pressure difference with the subscript \textit{``outlet,1''}, and the second outlet of the bifurcation, over which we are not currently computing the pressure difference, with the subscript \textit{``outlet,2''}.  In most ROMs, a pressure difference over a vascular junction is prescribed by the inclusion of an equation relating the pressure at the inlet branch to the pressure at each outlet branch.  In the context of \cref{fig:3d_to_0d}, this corresponds to setting the quantity $P_{\outl,1} - P_{\inl}$, which we hereafter refer to as $\Delta P$, to some value.  The equation
\begin{equation}
     \Delta P_{\text{static pressure}} = 0 
  \label{eq:constant_pressure}
\end{equation}
enforces conservation of static pressure.  Similarly, the equation
\begin{equation}
    \Delta P_{\text{total pressure}} = \frac{1}{2} \rho (u_{\inl}^2 - u_{\outl,1}^2)
  \label{eq:total_pressure}
\end{equation}
where $\rho$ is the density of blood enforces continuity of total pressure.

Aside from continuity of total and static pressure, the most commonly used model for pressure differences over vascular bifurcations is the Unified0D+ model, proposed in 2015 \cite{Mynard2015AJunctions, Chnafa2017ImprovedDrops, Mirramezani2020AFlow, Pewowaruk2021AcceleratedSwine, Blanco2018ComparisonReserve}.  It predicts the pressure difference over bifurcations as follows,
\begin{equation}
    \Delta P_{\text{Unified0D+}} = \left( 1-\frac{u_\inl}{u_{\outl,1}} \cos \left[ \frac{3}{4}(\pi - \theta ) \right] \right) \rho u_{\outl,1}^2,
    \label{eq:unified0d}
\end{equation}
where $u_\outl$ is the outlet velocity, $u_{\inl}$ is the velocity in the inlet branch, and $\theta$ is the angle between the outlet and inlet branch.  The Unified0D+ model incorporates physical principles such as conservation of mass, momentum, and energy along with empirically fitted corrections, but the absence of a term involving the time derivative of the flow limits the Unified0D+ model's ability to make accurate predictions on transient flow. A major contribution of \cite{Mynard2015AJunctions} was the introduction of a pseudodatum branch.  The pseudodatum is a modified inlet whose properties capture the effective behavior of multiple inlets and account for energy exchange between branches. Although not needed for the bifurcations considered in this work, the ability to accommodate junctions with arbitrary numbers of inlets and outlets is a major advantage of the Unified0D+ model. In \cref{sec:future_work}, we discuss a potential extension to our proposed model to handle these more complex junction types.

The Unified0D+ model predicts the pressure drop at the point at which the centerline bifurcates.  In contrast, our model predicts the pressure loss between the inlet and outlet of a bifurcation, some distance upstream and downstream of the bifurcation point.  We are interested in $\Delta P$ between the inlet and outlet as it encompasses the effects of the entire junction region, and it is the quantity needed in the ROM solvers we are considering.  To be able to compare the Unified 0D+ model to ours, we add the pressure differences expected in the vessel segment between the inlet and bifurcation point and between the bifurcation point and outlet to the pressure difference predicted by the Unified0D+ model.  The pressure differences in the inlet and outlet vessels are calculated assuming Poiseuille resistance as follows 
\begin{equation}
    \Delta P = \Delta P_{\text{Unified0D+}} + \Delta P_{\text{Poiseuille adjustment}} - \frac{8 \mu L_{\inl} Q_{\inl}}{\pi r_{\inl}^4} - \frac{8 \mu L_{\outl} Q_{\outl}}{{\pi r_{\outl}^4} },
    \label{eq:unified0d_poiseuille_correction}
\end{equation}
where $\mu$ is the viscosity of blood and $L$ is to the length of the vessel. $\Delta P_{\text{Poiseuille adjustment}}$ is a correction given in \cite{Mynard2015AJunctions} to be used when combining the Unified0D+ model with the Poiseuille equation-based vessel pressure difference model in this manner.  This system is similar to the approach taken in \cite{Qohar2021ASystem}.

We propose to model the pressure difference between the inlet and outlet of a bifurcation as a linear combination of the outlet flow $Q$, the square of the outlet flow $Q^2$, and the time derivative of the flow $\dot{Q}$ as follows
\begin{align}
    \Delta P_{\rri} = \reslm(\mathcal{G}) Q + \resqm(\mathcal{G}) Q^2 + L(\mathcal{G}) \dot{Q}.
    \label{lin_comb}
\end{align}
In the context of the circuit analogy, this formulation is equivalent to inserting a 0D block consisting of a serially connected resistor, quadratic resistor (similar to those used to represent stenosed vessels \cite{Mirramezani2019ReducedArteries, Lyras2021AnStenoses, Itu2013Non-invasiveMeasurements}), and inductor between the inlet and each outlet of a bifurcation.  For this reason, we hereafter refer to it as the Resistor-Resistor-Inductor (RRI) model.  The resistors characterized by $\reslm$ and $\resqm$ capture the steady behavior of flow through the bifurcation, while the inductor characterized by $L$ captures transient behavior.  In steady analyses, we refer to the (Resistor-Resistor) RR where there is no inductance ($L=0$).  Preliminary studies showed that the RRI model form was sufficient to closely replicate the relationship between the flow and pressure difference over a a wide range of bifurcation geometries in both steady and transient flows. 

\begin{figure}[htbp]
\centering
\includegraphics[width=0.4\textwidth]{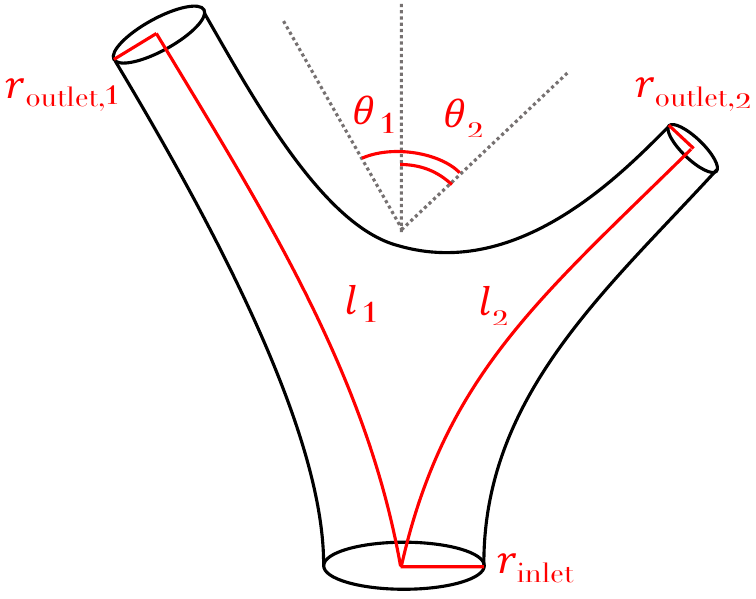}
\caption{Geometric parameters characterizing a bifurcation and used to predict the coefficients $\resqm$, $\reslm$, and $L$ which in turn govern the relationship between $Q$, $Q^2$, and $\dot{Q}$ and $\Delta P$ in the RRI model.}
\label{fig:labeled_geometry}
\end{figure}

We determined the coefficients, $\resqm$, $\reslm$, and $L$, from the bifurcation's geometry $\mathcal{G}$ using data-driven models.  The geometric features include $\mathcal{G}_1 = [r_\inl, r_{\outl, 1}, r_{\outl, 2}, \theta_1, \theta_2, l_1, l_2]$.  These are shown in \cref{fig:labeled_geometry} and refer to the inlet radius, outlet radius, auxiliary outlet radius, outlet angle, auxiliary outlet angle, outlet length, and auxiliary outlet length, respectively.  The auxiliary outlet is the bifurcation outlet for which we are not computing the pressure difference and is indicated with the subscript \textit{``2''}.

To train and validate the models, we generated synthetic bifurcation geometries and ran simulations for a series of flow conditions in each geometry to find the ground truth values of $\resqm$, $\reslm$, and $L$ associated with that geometry.

\begin{figure}[htbp]
\centering
\includegraphics[width=1\textwidth]{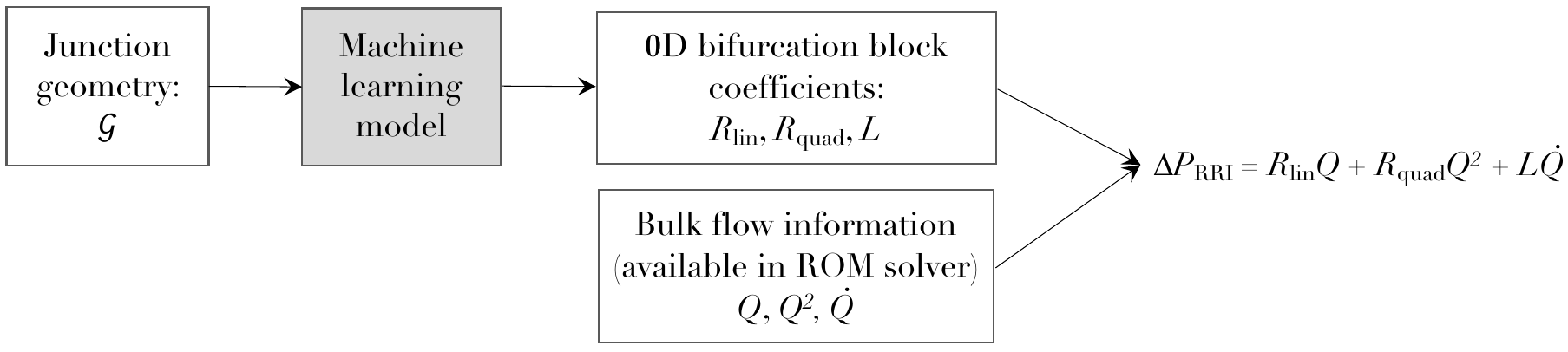}
\caption{Overview of the computation of a pressure difference over a vascular junction using the RRI model.}
\label{fig:RRI_flowchart}
\end{figure}

\subsection{Data Generation}
\label{sec:data_generation}
\begin{figure}[htbp]
\centering
\includegraphics[width=0.8\textwidth]{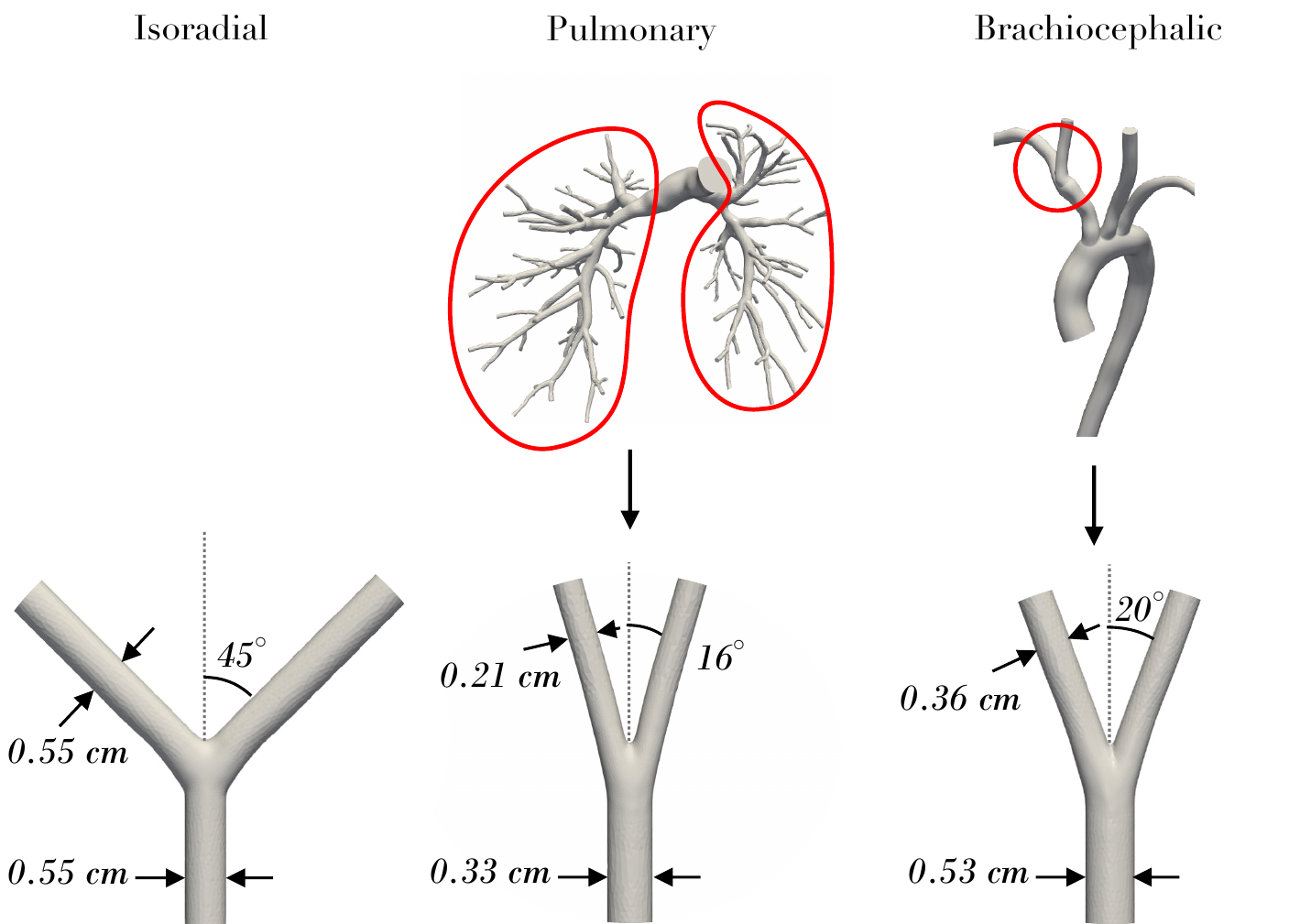}
\caption{Examples of pulmonary and brachiocephalic bifurcations in their surrounding vasculatures (top) and nominal idealized bifurcations from the isoradial, pulmonary, and brachiocephalic cohorts (bottom).}
\label{fig:junction_types}
\end{figure}

We generated three cohorts of idealized synthetic bifurcations representing: an isoradial cohort, a pulmonary cohort, and a brachiocephalic cohort. The isoradial cohort is representative of bifurcations analyzed in the work that led to the Unified0D+ model (although it should be noted that this study considered a wide range of outlet radii and offset angles) \cite{Mynard2015AJunctions}.  The pulmonary cohort is representative of distal bifurcations in the pulmonary tree.  The brachiocephalic cohort is representative of the bifurcation splitting the right subclavian artery and the right common carotid artery off of the brachiocephalic trunk.  These anatomies are shown in \cref{fig:junction_types}.  We consider the pulmonary and brachiocephalic cohorts to be the main benchmarks for our model, as they are based on bifurcations observed in the native vasculature.  Both the pulmonary and brachiocephalic cohorts represent anatomies for which junction pressure modeling is crucial---the brachiocephalic bifurcation can exhibit large magnitude pressure differences, and in pulmonary anatomies, many bifurcations are often chained together, so neglecting bifurcation pressure differences can result in significant cumulative error.

The parameters defining the automated generation of the bifurcation geometries were the inlet radius, the inlet and outlet radii, and the offset angles between the inlets and outlets. For bifurcations in the isoradial bifurcation cohort, these values were chosen randomly from a uniform distribution varying $\pm$ 20 \% of the parameter value for a nominal isoradial bifurcation. The characteristic geometric parameter values for bifurcations in the pulmonary and brachiocephalic cohorts were chosen randomly from uniform distributions spanning the 40th to 60th percentile of the range of parameter values observed in distal pulmonary and brachiocephalic bifurcations found in a publicly available database of patient-specific cardiovascular flow models, the Vascular Model Repository (VMR) \cite{Wilson2013TheResults} \footnote{\url{http://www.vascularmodel.com} (2022)}.  The ranges for the characteristic bifurcation dimensions can be found in \cref{app_parameter_ranges} of the appendix, and their average values are reported in \cref{fig:junction_types}.

We used SimVascular, an open-source software suite for cardiovascular modeling and simulation \cite{Updegrove2017SimVascular:Simulation}\footnote{\url{https://simvascular.github.io/} (May 2023)}, to create a set of idealized bifurcation solid models and simulate the flow fields associated with different boundary conditions.  In particular, we used the SimVascular Python API \footnote{\url{https://simvascular.github.io/documentation/python_interface.html}(May 2023)} to generate the geometries with automated scripts as follows.  First, we specified a series of points that define the centerlines of the vessels and identify the vessel lumen at each point.  Then, we lofted the vessels into solid models and merged them together to form a single geometry using Boolean operations.  A tetrahedral mesh was then generated from each geometry.  The mesh size was chosen to be the largest at which an accurate solution was attained, as determined by the mesh convergence studies shown in \Cref{mesh_convergence}.  The mesh was refined in the boundary layer as well as in a sphere surrounding the center of the junction to better resolve the complex flow behaviors in those locations.

Flows through the bifurcations were simulated using the stabilized finite element solver svSolver, provided with SimVascular, to solve the three-dimensional Navier Stokes equations.  A parabolic velocity profile with varying magnitude was prescribed at the inlet.  The inlet velocities for the isoradial cohort were sampled from a uniform distribution varying $\pm$ 20 \% of a nominal inlet velocity considered in \cite{Mynard2015AJunctions} while the inlet velocities applied to the pulmonary and brachiocephalic cohort were sampled from a uniform distribution ranging from 40th to 60th percentile of the inlet velocity values seen in pulmonary and brachiocephalic bifurcations in the VMR.  The inlet velocity ranges for each cohort are listed in \cref{app_parameter_ranges}.  Resistance boundary conditions were applied at the outlets with a fixed resistance value of 100 $\text{cm}^2 \text{s} \text{g}^{-1}$ and distal pressure of 0 mmHg \cite{Formaggia2010CardiovascularSystem}.

After the simulations were completed, we reduce the flow and pressure results to a 1D format by projection onto the model centerline.  To achieve this, at each point along the centerline we integrated the velocity field from the 3D flow results over the surface defined by the intersection of the 3D vessel with a plane normal to the centerline tangent vector and containing the centerline point.  Similarly, the pressure results were computed by calculating the average of the pressure field over the cross-section of the vessel normal to the centerline tangent.  The flow at the outlets and change in pressure with respect to the inlet were extracted from the 1D representation.  We defined inlet and outlet point locations to be about 10 inlet diameters downstream of the bifurcation point.  This distance was heuristically chosen, based on analysis of 3D flow in a range of bifurcations and flow conditions, to be large enough that we could assume the flow to be free of entrance effects and behave as fully-developed Poiseuille flow.  In this way, we ensured that all effects of the flow behaviors caused by the bifurcation will be analyzed and accounted for in our model.

\begin{figure}[htpb]

\includegraphics[scale=0.68]{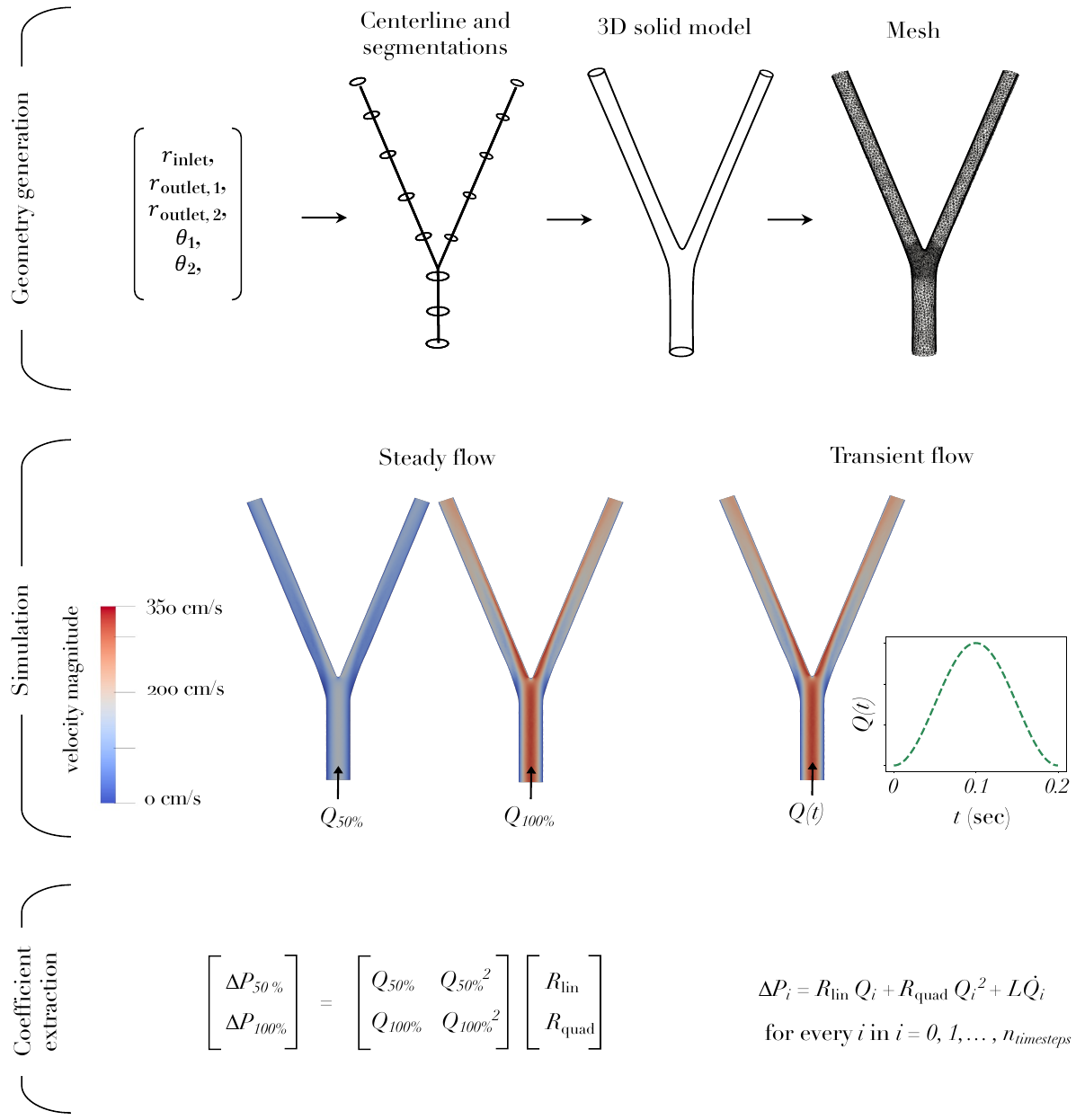}
\caption{Overview of data generation pipeline.  First, we generated a bifurcation geometry based on a set of geometric features.  Then, two steady simulations were run from which the coefficients $\reslm$ and $\resqm$ are determined by solving a simple system of equations containing the steady simulation results.  Finally, a transient simulation was run from which the coefficient $L$ is determined using least squares.  In the TO method, all three coefficients, $\reslm$, $\resqm$, and $L$ were determined from the transient simulation.}
\label{fig:data_gen_flow}
\end{figure}

For each geometry, two types of simulations were run---steady and transient.  First, two steady simulations were run at 50\% and 100\% of the sampled inlet flow rate.  A simulation was considered to have reached a steady state when the difference between the quantities of interest (outlet flow and pressure change) at the last time step and 100 time steps before the last time step was less than 1\% with a time step size of 0.001 seconds.  Second, a transient flow simulation was run in which the flow at the inlet was varied in time following the sinusoidal profile shown in \cref{fig:data_gen_flow}, where the maximum inlet flow rate was the sampled flow rate.  Using the simulation data, we found the coefficients $\reslm$, $\resqm$, and $L$ for each bifurcation geometry inlet-outlet pair.

First, we fit the coefficients $\reslm$ and $\resqm$ to the results of the steady simulations by solving the system of equations
\begin{equation}
    \begin{bmatrix}
    \Delta P_{50\%}\\
    \Delta P_{100\%}
    \end{bmatrix} = \begin{bmatrix}
    Q_{50\%} & Q_{50\%}^2\\
    Q_{100\%} & Q_{100\%}^2
    \end{bmatrix}\begin{bmatrix}
    \reslm\\
    \resqm
    \end{bmatrix}
  \label{eq:steady_state_lin_reg}
\end{equation}
where the subscripts 50\% and 100\% refer to the steady simulations run at 50\% and 100\% of the peak inlet flow.  We also experimented with running four steady simulations (adding $Q_{25\%}$ and $Q_{75\%}$) and fitting $\reslm$ and $\resqm$ using least squares.  We found that this did not result in significantly different values for $\reslm$ and $\resqm$, so to avoid the added computational cost, we proceeded with the two-simulation fitting method described in \ref{eq:steady_state_lin_reg}.

To determine the coefficients for the transient model, we used least squares on an over-defined system of equations containing the results of the transient simulations.  Each simulation timestep contributed a linear equation to the system of linear equations as follows
\begin{equation}
    \Delta P_i = \reslm Q_i + \resqm Q_i^2 + L \dot{Q}_i \qquad   \forall i = 0, 1, \dots, n_{\text{timesteps}}.
    \label{eq:transient_lin_reg}
\end{equation}
For transient flows, we considered two methods of fitting the coefficients.  In the first method, we substituted the values of $\reslm$ and $\resqm$ found from steady data corresponding to the same geometry into \cref{eq:transient_lin_reg}
and only fit $L$ from the transient data.  In the second method, we fit all three coefficients, $\reslm$, $\resqm$, and $L$ from the transient data.  We refer to the coefficients generated using the second method as transient-optimized (TO).

Having built a dataset of bifurcations, we next created ML models that take the bifurcation geometry as input and output the coefficients $\reslm$, $\resqm$, and $L$ that govern the relationship between $Q$, $Q^2$, and $\dot{Q}$ and $\Delta P$.  In particular, the ML models took as input a vector containing the geometric features of the bifurcation, $\mathcal{G}_1 = [r_\inl, r_{\outl,1}, r_{\outl,2}, \theta_1, \theta_2, l_1, l_2]$, and produced as output the vector containing the relevant coefficients.  In the standard modality, one model trained on resistance values fit to steady data outputted $\reslm$, $\resqm$, and a second model trained on inductance values fit to unsteady data outputted $L$.  In the TO method, a single ML model trained on coefficients fitted from unsteady data outputs all three coefficients, $\reslm$, $\resqm$, and $L$.  For each dataset, 80\% of the geometries were allocated to training the ML models and 20\% to testing them.  The training datasets included 149, 98, and 88 geometries, and the test sets include 38, 25, and 22 geometries for the isoradial, pulmonary, and brachiocephalic datasets, respectively. 

We tested several different ML model types, including K-Nearest Neighbors (KNN), Decision Trees (DT), Linear Regression (LR), Support Vector Regression (SVR), Gaussian Process Regression (GPR), and a Neural Network (NN) \cite{Kramer2013K-NearestNeighbors, Breiman2017ClassificationTrees, Awad2015SupportRegression, Rasmussen2005GaussianLearning, Aggarwal2023NeuralLearning, Hauck2014Scikit-LearnCookbook}.  With the exception of the Neural Network, each regression model was trained to minimize the squared difference between the coefficients predicted by the model and the coefficients extracted from the simulation data.  The NN was trained to minimize the squared difference between the pressure difference predicted by the coefficients given by the model and the pressure drop observed from simulation data.  This type of objective was implemented only for the NN because the backpropagation method of training a NN easily accommodates a customizable loss function. Such an approach would be much more difficult to implement for the other model types. Hyperparameter optimization for all models was conducted on the steady data from the brachiocephalic dataset using Ray Tune \cite{Liaw2018Tune:Training}, and the optimal parameters found are reported in \cref{hyperparameter_optimization}.

\section{Results}
\subsection{Steady Flows}

\begin{figure}[htbp]
\centering
\includegraphics[width = 1\linewidth]{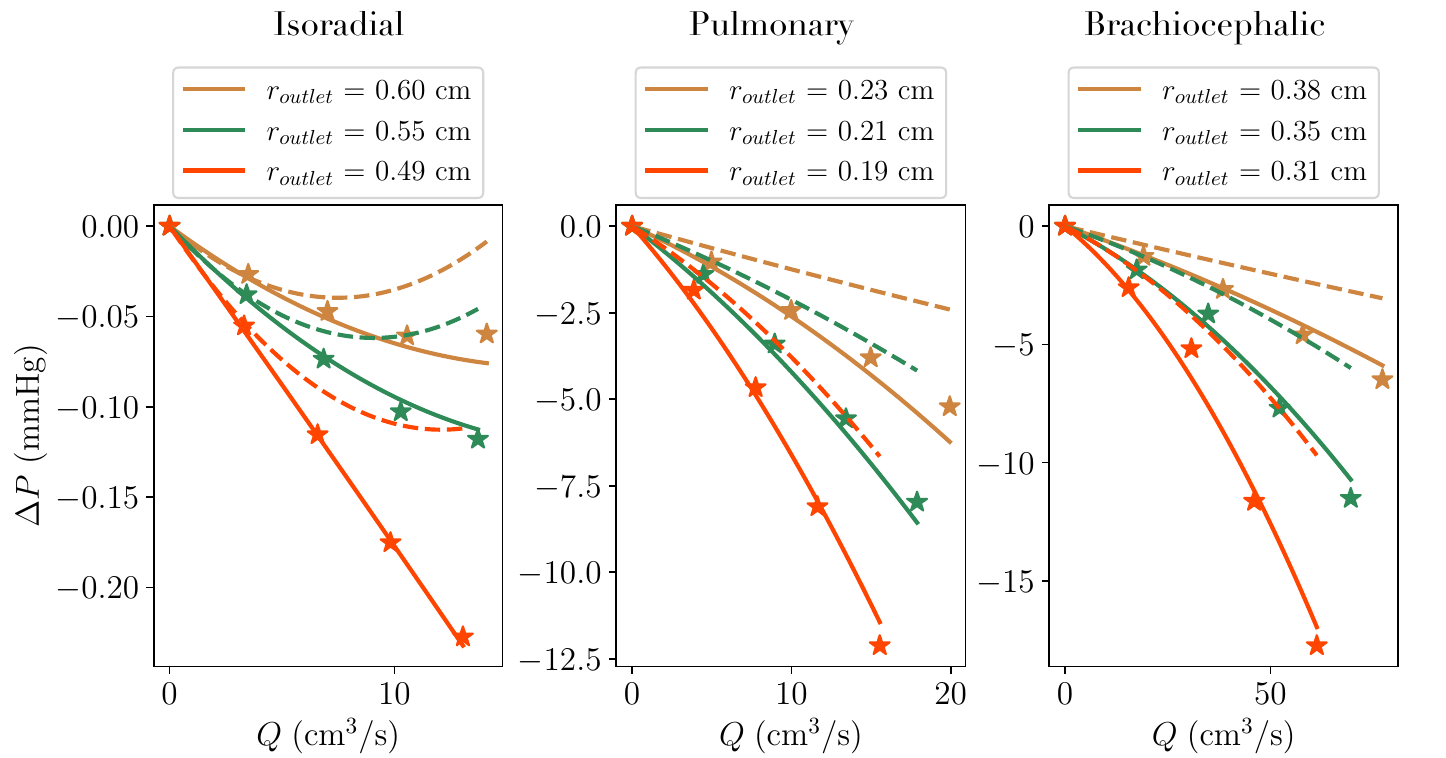}

\caption{Steady state $\Delta P$ vs $Q$ profiles for isoradial, pulmonary, and brachiocephalic type bifurcations.  Solid lines show the RR model prediction, dashed lines show the Unified0D+ model prediction, and stars show simulation results.  The different colors indicate different geometries, identical except for the outlet radius of one outlet vessel.}
\label{fig:steady_radius_sweep}
\end{figure}

In the first phase of our study, we analyzed steady flow through bifurcations.  In this case, $\dot{Q} = 0$, so there is no need to consider the inductance coefficient $L$.  We refer to the steady, inductance-free model as the RR model. We found overall that GPR and NN had the most success predicting the steady coefficients $\reslm$ and $\resqm$.  The accuracies of Unified0D+ model and our model using different regression techniques are compared in Table \ref{steady_aorta_table} for all three cohorts of bifurcations.  In \cref{fig:steady_radius_sweep}, we compare steady $\Delta P$-$Q$ profiles for predicted by 3D simulation, the RR model, and the Unified0D+ model for the three bifurcation types.  For each bifurcation type, we show three geometries, which differ only in the radius of the outlet vessel over which we are predicting the pressure difference.  For all cohorts, we see that the ground-truth 3D simulation predicts a lower pressure at the outlet than at the inlet, and that decreasing the outlet radius increases the magnitude of the pressure drop between outlet and inlet, as expected.  

We started by analyzing isoradial bifurcations similar to those considered in the original Unified 0D+ study.  We observed the $\Delta P$-$Q$ relationships predicted by the Unified0D+ model, our approach, and the CFD simulation results.  The simulated $\Delta P$-$Q$ relationships follow expected physical trends.  In isoradial bifurcations, the total outlet area is about double the total inlet area, so flow velocity is decreased in the outlet.  This deceleration causes a diminished dynamic pressure in the outlet, which contributes to elevated static pressure, resulting in a larger (less negative) $\Delta P$.  This effect becomes increasingly significant at higher flows, which cause more significant changes in dynamic pressure.  This is illustrated in the tendency of the $\Delta P$-$Q$ profiles to curve upwards.  We see that the curve is more pronounced for larger outlet radii, which again experience more significant changes in dynamic pressure.  While our model predicts the results of the CFD simulation quite closely, the Unified0D+ model consistently under-predicts the pressure drop.  The difference between the Unified0D+ model prediction and the CFD solution is more extreme at higher flow rates.  

Next, we performed a similar analysis on the pulmonary and brachiocephalic cohorts.  Unlike the isoradial bifurcations, these bifurcations have outlets with smaller radii than the inlet.  Since the total outlet area in these bifurcations is smaller than the inlet area, the flow accelerates upon entering the outlets, contributing to a decreased static pressure at the outlets.  This effect is more intense at higher flows and accounts for the concave-down shape of the $\Delta P$-$Q$ profiles.  Predictably, we see a more dramatic downward curve in the profiles for smaller outlet radius geometries.  The pulmonary bifurcations experienced flow rates similar to those of the isoradial cohort, but because the radii in the pulmonary cohort are much smaller, the pulmonary cohort exhibits higher velocities and larger pressure drops.  The brachiocephalic bifurcations have higher velocity flows than those in the other cohorts.  As expected, the magnitude of the pressure changes over these bifurcations is also larger.  Again, our model closely matches the CFD simulation results, but the Unified0D+ model significantly underestimates the magnitude of the pressure drop. 

\begin{table}[htbp]
    \centering
    \caption{Train and test root-mean-squared error in $\Delta P (Q)$ for steady flows in isoradial, pulmonary, and brachiocephalic type bifurcations using the steady RR model with different regression techniques.}
    \label{steady_aorta_table}    
    \begin{tabular}{l c c c c c c c c }
    \toprule
    Model type: & \multicolumn{6}{c}{RR} & Unified0D+ \\ 
    \cmidrule(r){2-7} \cmidrule(r){8-8} \\[-3pt]
    ML model type: & KNN & DT & LS & SVR & GPR & NN & \; \\

    \multicolumn{8}{c}{\;}\\ 
    \textit{Isoradial} & \multicolumn{7}{c}{\;}\\ 
    Train RMSE (mmHg)  & 0.016 & 0.014 & 0.022 & 0.0071 & 0.00074 & 0.0061 & \;\\
    Test RMSE (mmHg) & 0.022 & 0.018 & 0.020 & 0.012 & 0.013 & 0.012 & 0.10\\
    
    \multicolumn{8}{c}{\;}\\ 
    \textit{Pulmonary} & \multicolumn{7}{c}{\;}\\ 
    Train RMSE (mmHg)  & 0.63 & 0.47 & 0.93 & 0.95 & 0.018 & 0.41 & \;\\
    Test RMSE (mmHg) & 0.76 & 0.54 & 0.81 & 0.92 & 0.51 & 0.39 & 1.7\\

    \multicolumn{8}{c}{\;}\\ 
    \textit{Brachiocephalic} & \multicolumn{7}{c}{\;}\\ 
    Train RMSE (mmHg)  & 1.3 & 0.87 & 1.3 & 0.28 & 0.056 & 0.16 & \;\\
    Test RMSE (mmHg) & 1.5 & 1.07 & 1.3 & 0.78 & 0.33 & 0.34 & 4.9\\\bottomrule
    
    \end{tabular}
\end{table}

\subsection{Transient Flows}

\begin{figure}[htbp]
\centering
\includegraphics[width = 0.9\linewidth]{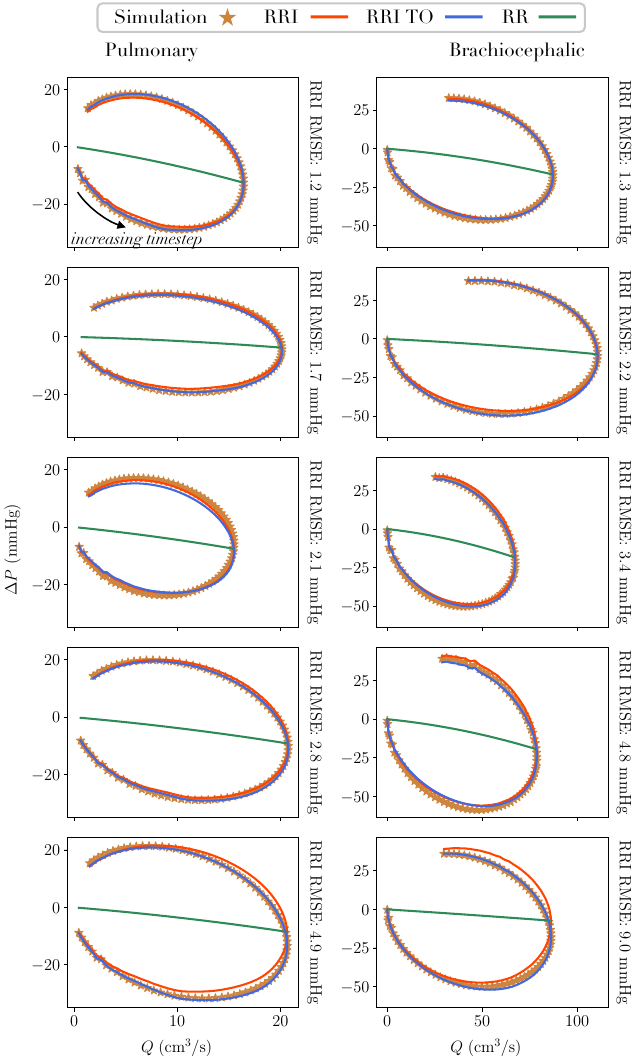}
\caption{Transient $\Delta P$ vs $Q$ profiles for test junctions from the pulmonary (left) and brachiocephalic (right) bifurcation cohorts. The sinusoidal flow profile shown in \cref{fig:data_gen_flow} was applied, where $Q$ starts at 0, increases, and then decreases, as indicated by the arrow in the top left panel. The geometries shown are those with the lowest, 25th percentile, median, 75th percentile, and highest RMSE for the standard RRI model out of the test set, from top to bottom. We show a comparison between our RRI model (standard and TO), the inductance-free RR model, and 3D simulation results.}
\label{fig:transient_flow_vs_predicted_dps}
\end{figure}

Next, we tested the performance of our RRI model on transient flows, which exhibit drastically different $\Delta P$-$Q$ profiles from steady-state flows (\cref{fig:transient_flow_vs_predicted_dps}).  For the same flow rate $Q$, the pressure difference over the bifurcation, $\Delta P$, can be radically different, depending on the derivative of the flow rate $Q$.  The inductor component of our model, $L \dot{Q}$ successfully captures this effect.  It is clear that without taking into account the time derivative of the flow, it is impossible to predict the pressure difference over the bifurcation with an acceptable accuracy.  To illustrate this, we include the inductance-free RR model in the visualization of our results.

In the standard RRI model, the coefficients $\reslm$, $\resqm$ are fit to match the steady simulation data, and $L$ is fit to the transient simulation data.  We also test the transient optimized (TO) model where all three coefficients, $\reslm$, $\resqm$, and $L$ are fit to match the transient simulation data.  As expected, the RRI TO model outperforms the standard RRI model. Again, we tested multiple regression techniques in our RRI model.  In the transient case, we found that, overall, the NN best predicted the coefficients $L$, $\reslm$, and $\resqm$ (\cref{transient_aorta_table}).

\begin{table}[htpb]
    \centering
\caption{Train and test root-mean-squared error in $\Delta P (Q)$ for transient flows in pulmonary-type and brachiocephalic-type bifurcations using the RRI model (standard and TO methods) with different regression techniques.}
\label{transient_aorta_table}
    \begin{tabular}{l c c c c c c c }
    \toprule
    ML model type: & KNN & DT & LS & SVR & GPR & NN & \; \\
     \multicolumn{7}{c}{\;}\\ 
     \textbf{Standard RRI model} & \multicolumn{6}{c}{\;}\\ 
    \multicolumn{7}{c}{\;}\\ 
    \textit{Pulmonary} & \multicolumn{6}{c}{\;}\\ 
    
    RRI Train RMSE (mmHg)  & 2.7 & 3.0 & 3.1 & 2.6 & 2.7 & 2.7\\
    RRI Test RMSE (mmHg) & 2.6 & 2.9 & 2.8 & 2.5 & 3.0 & 2.6\\

    \multicolumn{7}{c}{\;}\\ 
    \textit{Brachiocephalic} & \multicolumn{6}{c}{\;}\\ 
    
    RRI Train RMSE (mmHg)  & 4.0 & 4.3 & 4.5 & 3.8 & 3.8 & 4.0\\
    RRI Test RMSE (mmHg) & 4.3 & 5.3 & 5.2 & 4.5 & 4.5 & 4.6\\
     \multicolumn{7}{c}{\;}\\ 
     \textbf{Transient Optimized (TO) RRI model} & \multicolumn{6}{c}{\;}\\ 
     
    \multicolumn{7}{c}{\;}\\ 
    \textit{Pulmonary} & \multicolumn{6}{c}{\;}\\ 
    
    RRI Train RMSE (mmHg)  & 1.4 & 1.3 & 1.6 & 1.7 & 0.27 & 1.2\\
    RRI Test RMSE (mmHg) & 1.5 & 1.2 & 1.3 & 1.6 & 4.0 & 1.2\\

    \multicolumn{7}{c}{\;}\\ 
    \textit{Brachiocephalic} & \multicolumn{6}{c}{\;}\\ 
    
    RRI Train RMSE (mmHg)  & 2.8 & 2.3 & 2.4 & 1.2 & 0.80 & 1.3\\
    RRI Test RMSE (mmHg) & 3.5 & 3.1 & 3.1 & 2.4 & 2.6 & 2.2\\\bottomrule
    \end{tabular}
\end{table}

\section{Discussion}

We observe excellent performance of the RRI model on both steady and transient flows.  In steady flows, the proposed model outperforms the Unified0D+ model on all three bifurcation cohorts.  It is not surprising that the RRI model has higher accuracy than the Unified0D+ model on pulmonary and brachiocephalic bifurcations because these bifurcations have significantly different geometric features and experience much higher velocity flows than those considered in the development of the Unified0D+ model.  Furthermore, analysis of the 3D simulation results indicated that some of the assumptions used in the formulation of the Unified0D+ model, namely those about the velocity profile and distribution of total energy in the bifurcation may not be satisfied. Our model also accurately predicts pressure differences over vascular junctions in transient flows.  Inductors are traditionally used to capture inertial effects in the 0D electric circuit model, so it is expected that the inclusion of an inductor in the RRI model enables improved prediction of transient behavior that is impossible to account for without considering the time derivative of the flow \cite{Formaggia2010CardiovascularSystem}.  As expected, the TO model outperforms the standard RRI model, but the difference is very slight.  This indicates that the modeling of pressure differences can be cleanly split into a steady and transient component as done in the standard RRI model.  While marginally less accurate than the RRI TO model, the standard RRI model is more physically interpretable because $\reslm$ and $\resqm$ are found solely from steady data, which is consistent with the physical definition of resistors as circuit elements containing no time-dependent physics.  Furthermore, the standard RRI model is guaranteed to recover the steady-optimized coefficients when $\dot{Q} = 0$. The standard and TO model may each be preferable for different use cases, depending on the relative importance of interpretabilty versus accuracy.  Overall, we see the highest accuracy from the NN, GPR and SVR regression techniques.  We consider the NN to be most favorable because it demonstrated consistently high accuracy and because its structure integrates smoothly into our overall model's framework and can be easily generalized to handle more complex geometries in the future ({\cref{sec:data_generation}, \cref{sec:future_work}).

Our approach leverages physical knowledge to apply data-driven techniques in a constrained and judicious manner, providing an interpretable, robust, and practical method for predicting pressure differences over bifurcations.  First, formulating the bifurcation pressure loss model as a linear resistor, quadratic resistor, and inductor is physically intuitive and consistent with commonly used cardiovascular ROM approaches.  As such, it can be easily adopted by the community and implemented in existing ROM solvers.  Second, our formulation takes advantage of modern machine-learning techniques but mitigates the risk of unexpected behavior generally associated with ``black-box'' data-driven models.  Our model allows for ``sanity checks'' on the coefficients $\reslm$, $\resqm$, and $L$ (for instance, we expect $L$ to be negative). This type of interpretability is important for high-stakes applications like cardiovascular flow modeling.  Third, the proposed hybrid physics-ML formulation leverages a physics-based structure (instead of attempting to predict pressure differences directly from the bifurcation geometry and flow rate) which reduces the amount of data needed to train our models without overfitting.  Since generating training data is expensive in this application, this is a major advantage.

Finally, the structure of our model makes it straightforward to incorporate into existing ROM solvers.  When a vasculature containing a junction is represented as a ROM, the ML model is evaluated once to predict the $\resqm$, $\reslm$, and $L$ from the junction geometry.  Since these coefficients are only functions of the geometry, they remain fixed throughout the ROM simulation while the solution variables of $P$, $Q$, and $\dot{Q}$ vary as the solver iterates to converge to a solution that satisfies the system of equations characterizing the flow, as in \cref{fig:RRI_flowchart}. Computationally, this is advantageous from both an implementation and efficiency perspective. Notably, the ML model does not have to be evaluated at every solver iteration but only once as a preprocessing stage.  Furthermore, the derivative of the residual contribution from \eqref{lin_comb} with respect to the solution variables is computationally cheap to evaluate. 

\section{Limitations and Future Work}
\label{sec:future_work}

While this work takes significant steps towards improving ROMs by accounting for the pressure differences over vascular bifurcations, there are challenges to be overcome before the model we propose can be widely deployed.  Two major remaining challenges include the expansion of the model to cover a wider range of junctions and the incorporation of the model into ROM solvers.

To be most useful, our model should be able to predict the $\Delta P$-$Q$ relationship over any vascular junction it encounters.  In this work, we analyzed two idealized, limited cohorts of bifurcations that do not represent the full range of native vascular bifurcations.  To expand the applicability of our model, future studies should generate additional data so that the training dataset captures realistically observed variability in bifurcation geometries, including geometric complexities such as curvature, stenoses, and aneurysms. Furthermore, a junction pressure loss model may need to handle junction geometries more complex than simple bifurcations.  For instance, a vessel may split into more than two daughter branches or blood may flow from multiple inlets into one or more outlets (e.g., in backflow).  There are several approaches to handle this; for instance, future work could generalize our NN regressor, which predicts the coefficients that govern the $\Delta P$-$Q$ relationship between the inlet and outlet of a junction to a graph neural network (GNN) \cite{Scarselli2009TheModel}.  Thanks to its flexible structure, a GNN will be capable of handling junctions with any number of inlets and outlets.  Having seen in this work that NNs are capable of capturing the behavior of flow in a bifurcation; we predict that a GNN will be able to predict the $\Delta P$-$Q$ relationship on a junction with an arbitrary number of inlets and outlets with similar accuracy.

A second challenge will be the incorporation of our RRI model into current ROMs.  In most cases, this will require only minimal changes to the code.  For instance, in SimVascular's 0D solver, the only necessary change will be to replace the equations enforcing equal pressure between junction outlets and inlets with equations enforcing the pressure difference predicted by our model.  The greater difficulty will be the standardization of the definition of a junction.  This is a general challenge encountered when translating 3D vasculatures into reduced-order systems because there is no straightforward definition of the boundary between a junction and a vessel.  In this work, we defined the vessel-bifurcation boundaries heuristically, based on where the flow exhibits fully developed Poiseuille behavior, free of splitting effects caused by the bifurcation.  This system worked for our study because we had a standardized geometry-generation method and an understanding of the flow-splitting behavior, but it would not work for generic use where there is no a priori general knowledge of the flow.  The inaccuracies introduced by uncertainty in junction definition may be somewhat mitigated by the inclusion of the junction length in the feature set supplied to the ML model, but complicate the problem and may present difficulties when the RRI model encounters alternatively defined junctions. These challenges should be addressed in future studies that demonstrate improved performance of ROM solvers with the RRI model added.  

\section{Conclusion}

We presented an RRI model which models the pressure difference between a bifurcation inlet and outlet as the voltage difference over a serially connected linear resistor, quadratic resistor, and inductor and uses ML to predict the resistances and inductance from the bifurcation geometry. To generate data on which to train and validate our model, we developed an automated pipeline to generate and simulate flow through bifurcations.  We found that our RRI model performed well on various geometry types and in both steady and transient flows.  As such, it presents a viable method to account for bifurcation pressure differences in cardiovascular ROMs, thus overcoming a major limitation of their accuracy.  More accurate ROMs will add significant value to the cardiovascular flow modeling community, both in support of 3D CFD simulations as surrogate models for many-query applications (e.g., uncertainty quantification and boundary condition tuning) and in their own right as stand-alone models for real-time applications.  In addition to contributing to more accessible and accurate patient-specific cardiovascular flow analysis, this work highlights opportunities for synergy between 3D CFD, physics-based reduced-order modeling, and data-driven techniques in medical research.

\section*{Acknowledgments}

This work was funded by the Stanford Graduate Fellowship and the National Science Foundation Graduate Research Fellowship Program.  Additional support was provided by NIH Grants R01LM01312003, R01EB02936204,  R01HL16751601, and K99HL161313, and the Stanford Maternal and Child Health Research Institute.  The authors also thank Dr.\ Karthik Menon and Zachary Sexton for their helpful insight and support.

\bibliographystyle{unsrtnat}
\bibliography{references}

\section{Appendix}

\newpage
\subsection{Mesh Convergence}
\label{mesh_convergence}
Studies showing the convergence of steady flow simulations for decreasing mesh size.  These studies guided our choice of mesh size.
\begin{figure}[htbp]
\centering
\includegraphics[scale= 0.6]{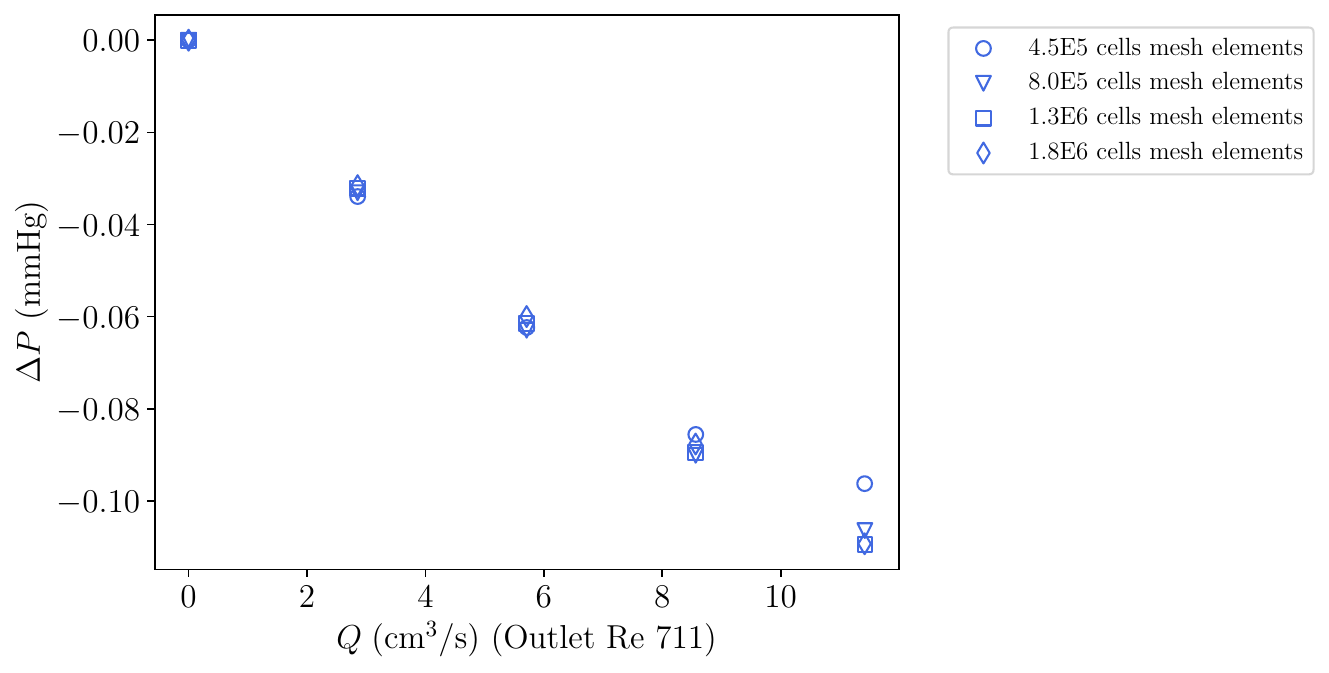}
\caption{Mesh refinement on isoradial bifurcation.}
\label{fig:mesh_ref_mynard}
\end{figure}

\begin{figure}[htbp]
\centering
\includegraphics[scale= 0.6]{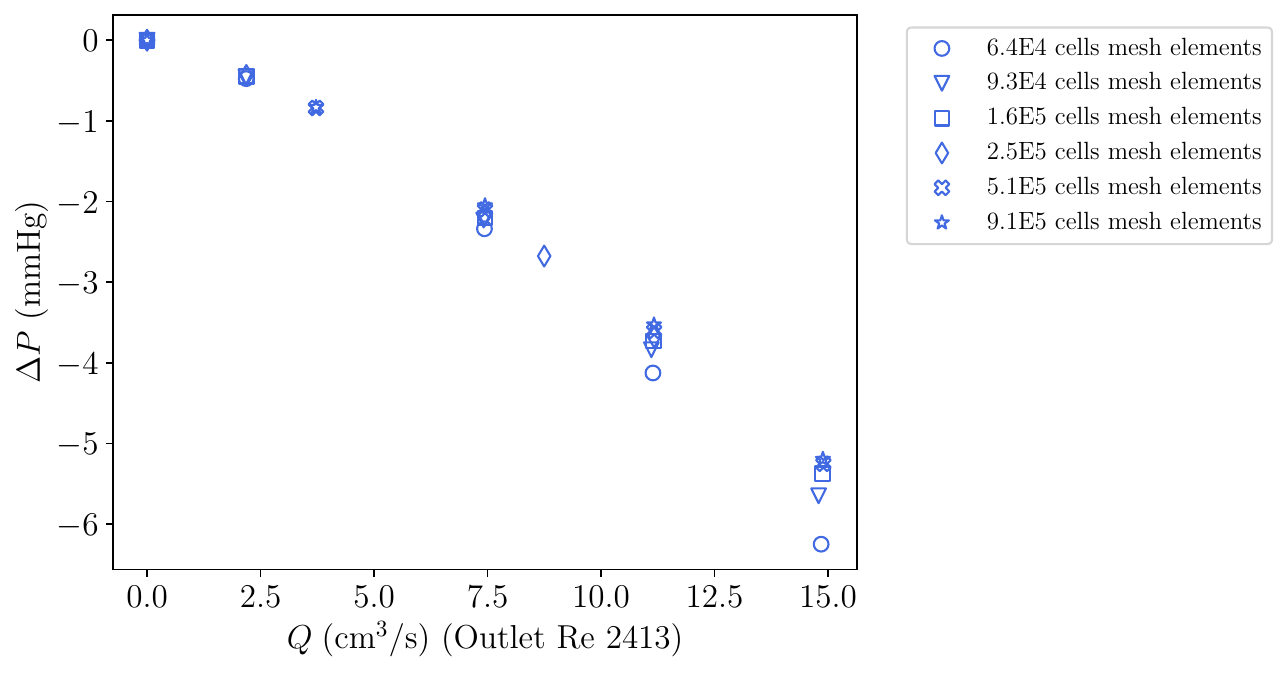}
\caption{Mesh refinement on pulmonary bifurcation.}
\label{fig:mesh_ref_pulmo}
\end{figure}

\begin{figure}[htbp]
\centering
\includegraphics[scale= 0.6]{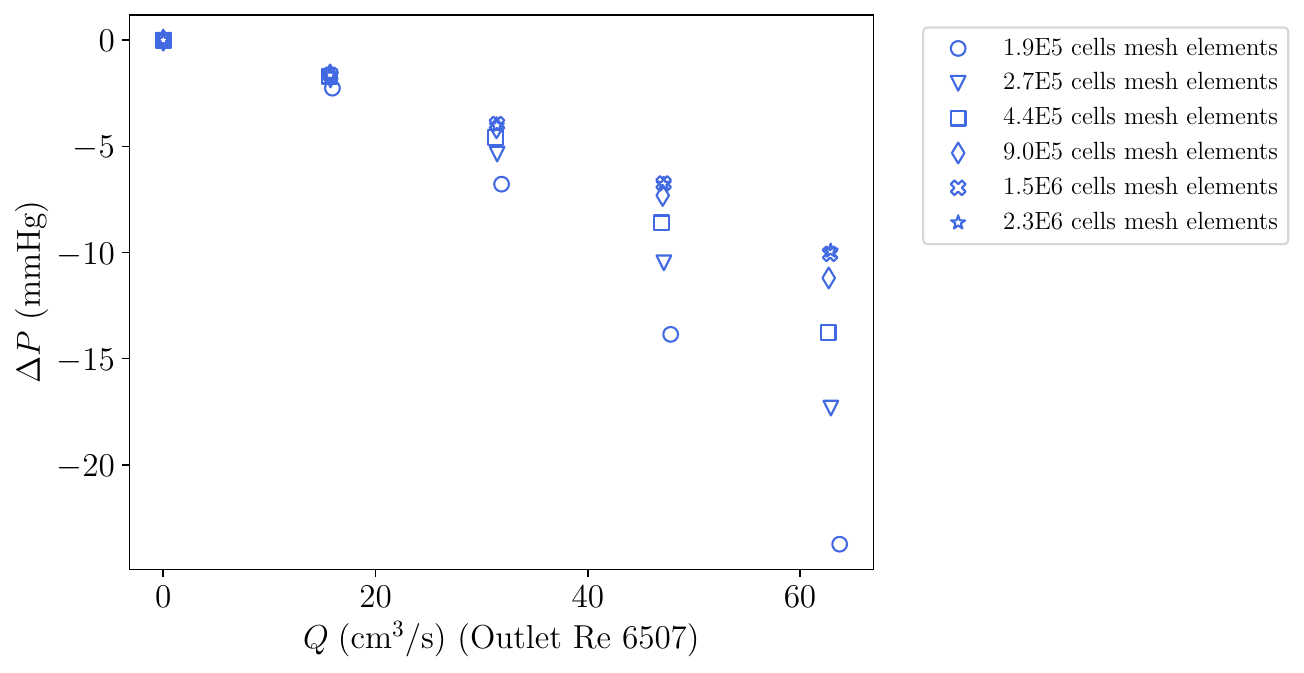}
\caption{Mesh refinement on brachiocephalic bifurcation.}
\label{fig:mesh_ref_aorta}
\end{figure}

\newpage
\subsection{Parameter Ranges}
\label{app_parameter_ranges}
Ranges of values from which the geometric parameters characterizing the isoradial and brachiocephalic cohorts of bifurcations were  uniformly sampled.  The isoradial ranges were found by varying the nominal parameters used in \cite{Mynard2015AJunctions} $\pm 20\%$.  The pulmonary and brachiocephalic ranges were the 40th to 60th percentiles of the range of parameter values observed in distal pulmonary and brachiocephalic bifurcations, respectively, found in the VMR \cite{Wilson2013TheResults}.  \textit{Note:} for the pulmonary and brachiocephalic cohorts, the outlet radii were not sampled, but were computed as the product of the inlet radius and outlet-inlet radius ratio, both of which were sampled.
\begin{table}[htpb]
    \centering
    \begin{tabular}{l r c c c c}
    \textit{Parameter}: & Inlet Radius (cm) & Outlet Radius (cm) & Outlet Angle ($^\circ$) & Inlet Velocity (cm/s) \\
     \cmidrule(r){2-2} \cmidrule(r){3-3} \cmidrule(r){4-4} \cmidrule(r){5-5} 
    Isoradial  & 0.44 - 0.66 & 0.44 - 0.66 &36 - 54 & 49 - 74\\
    Pulmonary & 0.28 - 0.37 & 0.16 - 0.27 & 13 - 19 & 95 - 140 \\
    Brachiocephalic & 0.46 - 0.59 & 0.28 - 0.43 & 16 - 24 & 127 - 180 \\
    \end{tabular}
    \caption{Ranges for parameters in isoradial, pulmonary, and brachiocephalic bifurcation datasets.}
    \label{parameter_ranges}
\end{table}

\subsection{Hyperparameter Optimization}
Hyperparameters chosen using Ray Tune for the candidate ML models \cite{Liaw2018Tune:Training}.  The objective being minimized with Ray Tune was the error on the test set of steady flows through brachiocephalic junctions, and the resulting hyperparameters were used for all 3 cohorts for steady and transient models. \textit{ Note:} for the NN trained on the isoradial cohort, a hidden layer size of 70 was found (manually) to be optimal.
\label{hyperparameter_optimization}

\begin{table}[htpb]
    \centering
    \begin{tabular}{ l r c c c c c c}
    Parameter & Value\\
    \cmidrule(r){1-2}
    \multicolumn{2}{c}{\textit{K Nearest Neighbors}}\\
    \cmidrule(r){1-2}
    number of neighbors  & 7 \\
    \cmidrule(r){1-2}
    \multicolumn{2}{c}{\textit{Decision Tree}}\\
    \cmidrule(r){1-2}
    maximum depth  & 4 \\
    minimum samples per leaf & 8 \\
    \cmidrule(r){1-2}
    \multicolumn{2}{c}{\textit{Support Vector Regression}}\\
    \cmidrule(r){1-2}
    $C$ (L2 regularization parameter)  & 1.4 \\
    $\epsilon$ (no-penalty margin) & 0.029 \\
    \cmidrule(r){1-2}
    \multicolumn{2}{c}{\textit{Gaussian Process Regression}}\\
    \cmidrule(r){1-2}
    $\alpha$ (regularization parameter)  & 0.0020 \\
    radial basis function kernel length scale & 1.6 \\
    \cmidrule(r){1-2}
    \multicolumn{2}{c}{\textit{Neural Network}}\\
    \cmidrule(r){1-2}
    hidden layer size & 48 \\
    number of hidden layers & 2 \\
    learning rate  & 0.018 \\
    learning rate decay & 0.031 \\
    batch size & 24 \\

    \end{tabular}

    \caption{Values found using Ray Tune for hyperparameter optimization of candidate ML models.}
    \label{HPO_values}
\end{table}

\subsection{Verification of Correct Unified0D+ Model Implementation}
Studies replicating plots in \cite{Mynard2015AJunctions} to confirm that our implementation of the Unified 0D+ model was consistent.

\begin{figure}[htbp]
\centering
\includegraphics[scale= 0.5]{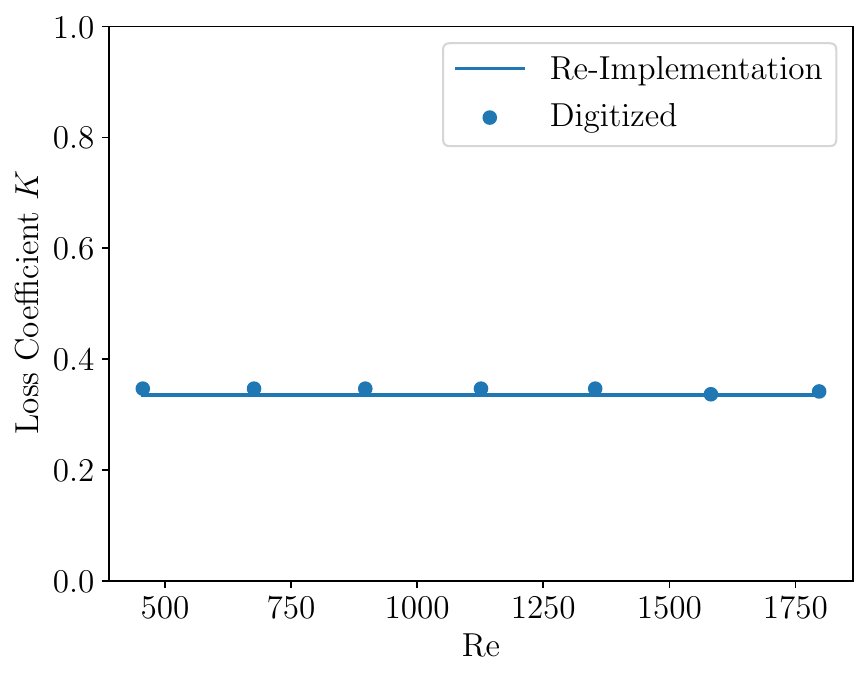}
\includegraphics[scale= 0.5]{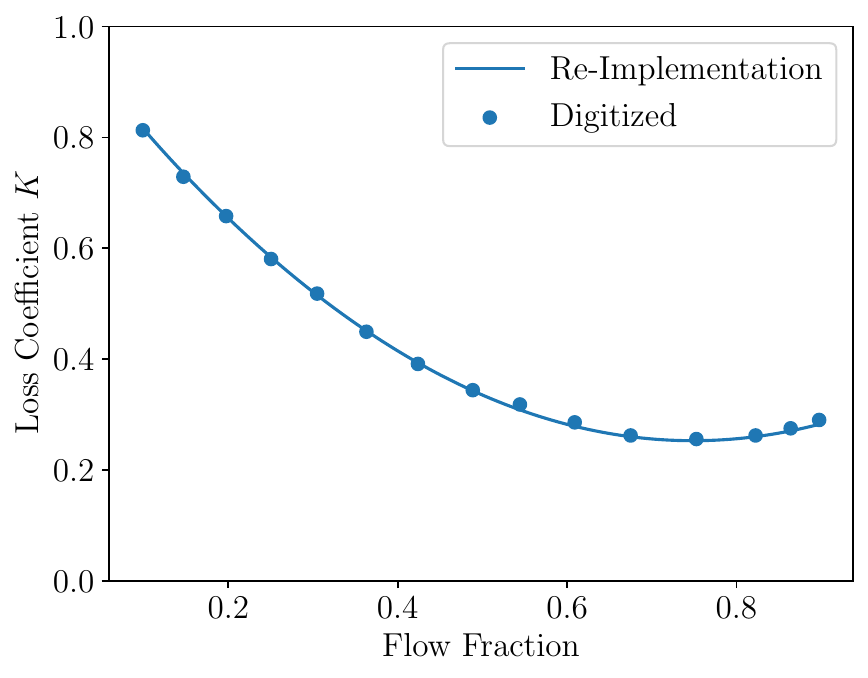}
\includegraphics[scale= 0.5]{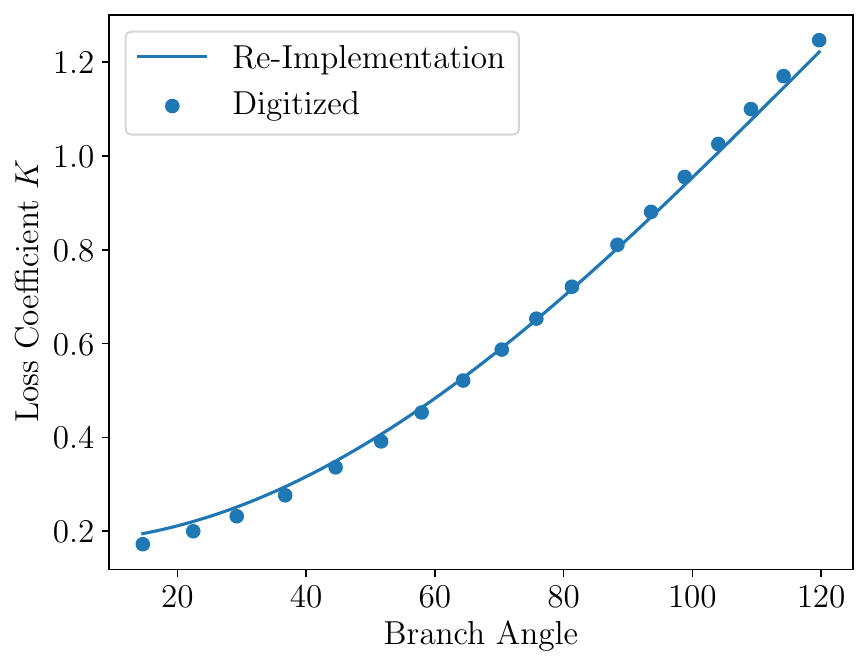}
\vspace{-5 pt}
\caption{Replicating Figure 8a, 8b, and 8c (top to bottom) of \cite{Mynard2015AJunctions}.}
\label{fig:mynard_fig_8c_comp}
\end{figure}

\end{document}